\documentclass[12pt]{iopart}
\usepackage{iopams}
\usepackage{array}
\usepackage{graphicx}
\usepackage{caption}
\usepackage{subcaption}
\usepackage{hyperref}
\usepackage{hypernat}
\usepackage{color}

\begin{document}
\title[A measure of dissimilarity between diffusive processes on networks]{A measure of dissimilarity between diffusive processes on networks}
\author{Alejandro P. Riascos${}^{1}$ and Francisco Hern\'andez Padilla${}^{2}$}  
\address{${}^1$Instituto de F\'isica, Universidad Nacional Aut\'onoma de M\'exico, 
	C.P. 04510, Ciudad de M\'exico, M\'exico\\
	${}^2$ Departamento de F\'isica, Facultad de Ciencias, Universidad Nacional Aut\'onoma de M\'exico, Ciudad Universitaria,  Ciudad de M\'exico 04510, M\'exico}

\begin{abstract}
In this paper, we present a framework to compare the differences in the occupation probabilities of two random walk processes, which can be generated by modifications of the network or the transition probabilities between the nodes of the same network. We explore a dissimilarity measure defined in terms of the eigenvalues and eigenvectors of the normalized Laplacian of each process. This formalism is implemented to examine differences in the diffusive dynamics described by circulant matrices, the effect of new edges, and the rewiring in networks as well as to evaluate divergences in the transport in degree-biased random walks and random walks with stochastic reset. Our results provide a general tool to compare dynamical processes on networks considering the evolution of states and capturing the complexity of these structures.
\end{abstract}


\maketitle
\section{Introduction}
Random walks are present at different scales in nature and find applications to a broad range of fields. In particular, stochastic trajectories represented as a succession of discrete steps naturally describe varied processes like diffusion, chemical reactions, animal movements, and search processes in general. In several applications in the study of complex systems it is convenient to explore all these dynamical processes when the activity takes place on a network \cite{VespiBook}. Random walks that move between nodes in a network are relevant to many problems and constitute the natural framework to study diffusive processes in regular and irregular structures \cite{VespiBook,Hughes,Lovasz1996,BlanchardBook2011}. Network exploration by random walks can be defined through hops to nearest neighbors \cite{ReviewJCN_2021,NohRieger2004,MasudaPhysRep2017} or with long-range jumps between distant nodes \cite{ReviewJCN_2021,RiascosMateos2012,RiascosMateosFD2014,FractionalBook2019}. The understanding of the relation between the random walk dynamics and the network topology requires a particular treatment in terms of matrices and spectral methods \cite{NewmanBook,GodsilBook, VanMieghem2011}.
\\[2mm]
The complexity present in networks and the different dynamical processes that may occur in these structures motivates the exploration of measures to quantify the differences between two dynamics; for example, to characterize the effect of a modification in the network's connectivity or to evaluate how changes in the strategy followed by a random walker affect the exploration of the structure.  Recent findings include distances between networks that usually fall in one of two general categories defining structural and spectral distances, often considered mutually exclusive \cite{Donnat_2018}. The first one captures variations in the local structure, as examples of this metric are the Hamming \cite{Hamming1950} and, Jaccard distances \cite{Jaccard1901,LEVANDOWSKY1971} characterizing the number of edge deletions and insertions necessary to transform one network into another. In contrast, the spectral approach assesses the smoothness of the evolution of the overall structure by tracking changes in functions of the eigenvalues of the graph Laplacian, the normalized Laplacian or simply the adjacency matrix \cite{Donnat_2018}. As examples of spectral measures,  we have the spanning-tree similarity \cite{Donnat_2018} and several distances based on the eigenspectrum distributions \cite{Donnat_2018,Hartle2020}. A different possible perspective focuses on the use of graph kernels to define similarities between graphs \cite{Jurman2015,Hammond2013,Scott2021}.
\\[2mm]
In this contribution, we study diffusive processes over a network, modeled using a random walk. We present a framework to compare the diffusion on networks in terms of the evolution of states. The measure introduced is defined using the eigenvalues and eigenvectors of the normalized Laplacian of each process to examine differences generated by modifications in the networks or the dynamics. The paper is organized as follows: In Sec. \ref{Sec_General_Theory} we introduce different quantities to describe the diffusion on networks; in particular, the normalized Laplacian of a graph $\hat{\mathcal{L}}$ that defines the process and the continuous-time evolution of states describing the diffusion. We also introduce a general definition for the dissimilarity of states to compare a dynamical process defined by $\hat{\mathcal{L}}$ and a modified dynamics described by $\hat{\mathcal{L}}^\prime$, this second matrix describes a process with modifications in the connectivity of the network or changes in the transition probabilities that define the random walker. In Sec. \ref{Section_Circulant}, we apply the general formalism to cases where  $\hat{\mathcal{L}}$ and  $\hat{\mathcal{L}}^\prime$ are circulant matrices. We explore analytically the evolution of the dissimilarity between states for random walks with bias on rings and the effect of adding weighted edges in interacting cycles. In Sec. \ref{Sec_rewiring}, we illustrate the consequences of rewiring a network. Mainly, the effect of an additional link in a ring and the stochastic reorganization of links in the Watts-Strogatz model.  In Sec. \ref{Sec_DifferentRWs}, we examine processes for which the network is the same but considering modifications in the definition of the random walker. We explore degree-biased random walks \cite{ReviewJCN_2021,FronczakPRE2009} and dynamics with stochastic reset to the initial node \cite{ResetNetworks_PRE2020,Reset_2022}. This last case is explored analytically to measure the effect of reset when compared with the case without restart in different types of structures including deterministic and random networks. Our findings show how the measures implemented allow for quantifying differences in diffusive dynamics. The methods developed in this research are general and open the doors to a broad spectrum of tools applicable to different random walk strategies and dynamical processes on networks.
\section{General theory}
\label{Sec_General_Theory}
In this section, we present general definitions for diffusion on networks and introduce a dissimilarity measure to compare two diffusive processes on these structures. 
\subsection{Diffusion on networks}
\label{Sec_General_Theory_Diffusion}
We consider connected networks with $N$ nodes $i=1,\ldots,N$ described by the adjacency matrix $\mathbf{A}$ with elements $A_{ij}=1$ if there is a link between the nodes $i$, $j$, and $A_{ij}=0$ otherwise; also $A_{ii}=0$ to avoid self-loops in the network. The degree of the node $i$ is given by $k_i=\sum_{l=1}^N A_{il}$.
\\[2mm]
In addition, we have a Markovian random walk on the network where at
each time  $t=0,\Delta t,2\Delta t,\ldots$ the random walker hops from node $i$ to node $j$ with a transition probability $w_{i\to j}$ in a process without memory of the visited sites. All the elements  $w_{i\to j}$ define the dynamics on the network through a transition matrix $\mathbf{W}$ \cite{Hughes,NohRieger2004}, in general, this is a stochastic matrix due to the condition $\sum_{j=1}^N w_{i\to j}=1$.
\\[2mm]
The occupation probability  $p_{ij}(t)$ to start at time $t=0$ on node $i$ and to reach the node $j$ at time $t$ satisfies the master equation \cite{Hughes,NohRieger2004,FractionalBook2019}
\begin{equation}
p_{ij}(t+\Delta t) = \sum_{l=1}^N  p_{il} (t) w_{l\rightarrow j}.
\end{equation}
For small $\Delta t$ the following approximation is valid
\begin{eqnarray}
\frac{d p_{ij}(t) }{dt}\approx-\frac{1}{\Delta t}\left[p_{ij} (t)-\sum_{l=1}^N  p_{il} (t) w_{l\rightarrow j}\right]=-\frac{1}{\Delta t}\sum_{l=1}^Np_{il} (t)\left[\delta_{lj}-   w_{l\rightarrow j}\right],
\end{eqnarray}
where $\delta_{ij}$ denotes the Kronecker delta. In this manner, for continuous-time, the dynamics of the random walker is defined in terms of the normalized Laplacian matrix $\hat{\mathcal{L}}$ with elements $\mathcal{L}_{ij}=\delta_{ij}-w_{i\to j}$. Therefore, we have the master equation \cite{VespiBook}:
\begin{equation} \label{master}
\frac{d p_{ij}(t)}{dt}=-\sum_{l=1}^N\,p_{il}(t)\mathcal{L}_{lj}\, ,
\end{equation}
where we express the time $t$ in units $\Delta t$. Integrating Eq. (\ref{master}), all the temporal evolution of $p_{ij}(t)$ is determined by the operator $\hat{\mathcal{U}}(t)=\exp[-\hat{\mathcal{L}}\,t]$. Using Dirac's notation, $p_{ij}(t)$ is given by
\begin{equation}
p_{ij}(t)=\langle i | \exp[-\hat{\mathcal{L}}\,t]|j\rangle.
\end{equation}
Here $|i\rangle$ denotes the vector with all its components equal to 0 except the $i$-th one, which is equal to 1, $\langle i|=|i\rangle^\mathrm{T}$ where  $\mathrm{T}$ denotes the transpose of vectors.
\\[2mm]
Let us now introduce the state $\langle \bar{\Psi}_i(t)|$ that evolves with $\hat{\mathcal{U}}(t)$ and is given by
\begin{equation}\label{stateLleft}
\langle \bar{\Psi}_i(t)|\equiv\langle i | \exp[-\hat{\mathcal{L}}\,t].
\end{equation}
Similarly, we have a second process that occurs on a network with $N$ nodes. The process is described by the matrix $\hat{\mathcal{L}}^\prime$. The differences between $\hat{\mathcal{L}}$ and $\hat{\mathcal{L}}^\prime$ can be caused by network alterations as for example the addition of one or multiple edges or by modifications in the strategy that each random walker follows. The temporal evolution of the second process is given by
\begin{equation}\label{stateLprimeleft}
\langle \bar{\Psi}_i^\prime(t)|\equiv\langle i | \exp[-\hat{\mathcal{L}}^\prime\,t].
\end{equation}
The state vectors $\langle \bar{\Psi}_i(t)|$ and $\langle \bar{\Psi}_i^\prime(t)|$ have the counterpart as column vectors
\begin{equation}\label{statesright}
 |\Psi_i(t)\rangle \equiv \exp[-\hat{\mathcal{L}}\,t]|i\rangle,\qquad |\Psi_i^\prime(t)\rangle \equiv \exp[-\hat{\mathcal{L}}^\prime\,t]|i\rangle. 
\end{equation}

\subsection{A measure of dissimilarity}
\label{Subsec_Ds}
In order to establish a ``dissimilarity'' between the two dynamical processes defined by the modified Laplacians $\hat{\mathcal{L}}$ and $\hat{\mathcal{L}}^\prime$,  we apply the equation for the cosine dissimilarity to compare the states of the systems generated for each operator and given by
\begin{equation}\label{DissimilarM_nodei}
\mathcal{D}_i(\hat{\mathcal{L}},\hat{\mathcal{L}}^\prime;t)\equiv 1-\frac{\langle \bar{\Psi}_i(t)| \Psi_i^\prime(t) \rangle}{\sqrt{\left|\langle \bar{\Psi}_i(t)| {\Psi}_i(t) \rangle\right| \left|\langle \bar{\Psi}_i^\prime(t)| \Psi_i^\prime(t) \rangle\right|}}.
\end{equation}
Although $\mathcal{D}_i(\hat{\mathcal{L}},\hat{\mathcal{L}}^\prime;t)$ measures the ``dissimilarity'' between two processes, it is not a distance because it 
is not a metric measure.  In this respect, it is worth noticing that in the general case in  Eq. (\ref{DissimilarM_nodei}), we have
\begin{eqnarray}\nonumber
\langle \bar{\Psi}_i(t)| \Psi_i^\prime(t)\rangle&=\langle i| \exp[-\hat{\mathcal{L}}\,t]\exp[-\hat{\mathcal{L}}^\prime\,t]|i \rangle \\
&\neq \langle i| \exp[-\hat{\mathcal{L}}^\prime\,t]\exp[-\hat{\mathcal{L}}\,t]|i \rangle= \langle \bar{\Psi}^\prime_i(t)| \Psi_i(t) \rangle.
\end{eqnarray}
Therefore $\langle \bar{\Psi}_i(t)| \Psi_i^\prime(t)\rangle\neq \langle \bar{\Psi}^\prime_i(t)| \Psi_i(t) \rangle$, 
except when $\hat{\mathcal{L}}$ and $\hat{\mathcal{L}}^\prime$ commute. As a consequence    $\mathcal{D}_i(\hat{\mathcal{L}},\hat{\mathcal{L}}^\prime;t)\neq \mathcal{D}_i(\hat{\mathcal{L}}^\prime,\hat{\mathcal{L}};t)$, making important in the application of Eq. (\ref{DissimilarM_nodei}) the order of the original process used as a reference $\hat{\mathcal{L}}$ and the modified dynamics $\hat{\mathcal{L}}^\prime$. The definition in Eq. (\ref{DissimilarM_nodei}) is motivated by the cosine similarity of vectors in a Euclidean space defined in terms of the dot product of vectors, the angle between the vectors is a measure of the similarity. In the comparison of states describing diffusive dynamics, $\mathcal{D}_i(\hat{\mathcal{L}},\hat{\mathcal{L}};t)=0$ and $\mathcal{D}_i(\hat{\mathcal{L}},\hat{\mathcal{L}}^\prime;t)\in \mathbb{R}$. An analogous definition of the dissimilarity to compare the effect of different initial conditions in the dynamics of a random walker has been recently explored by Ghavasieh et. al. in Ref. \cite{DomenicoPRE2020}.
\\[2mm]
On the other hand, the evaluation of Eq. (\ref{DissimilarM_nodei}) requires the spectral decomposition of the matrices $\hat{\mathcal{L}}$, $\hat{\mathcal{L}}^\prime$. For ergodic random walks, the transition matrix $\mathbf{W}$ can be diagonalized. For right eigenvectors of $\mathbf{W}$ we have $\mathbf{W}\left|\phi_i\right\rangle=\lambda_i\left|\phi_i\right\rangle $ for $i=1,\ldots,N$, where for the set of eigenvalues $\lambda_1=1$ is unique and $|\lambda_m|\leq 1$ for $m=2,\ldots,N$. On the other hand, from right eigenvectors we define a matrix $\mathbf{Z}$ with elements $Z_{ij}=\left\langle i|\phi_j\right\rangle$. The matrix $\mathbf{Z}$ is invertible, and a new set of vectors $\left\langle \bar{\phi}_i\right|$ is obtained by means of $(\mathbf{Z}^{-1})_{ij}=\left\langle \bar{\phi}_i |j\right\rangle $, then
\begin{equation}\label{cond1}
\delta_{ij}=(\mathbf{Z}^{-1}\mathbf{Z})_{ij}=\sum_{l=1}^N \left\langle\bar{\phi}_i|l\right\rangle \left\langle l|\phi_j\right\rangle=\langle\bar{\phi}_i|\phi_j\rangle \,
\end{equation}
and 
\begin{equation}\label{cond2}
\mathbb{I}=\mathbf{Z}\mathbf{Z}^{-1}=\sum_{l=1}^N \left|\phi_l\right\rangle \left\langle \bar{\phi}_l \right| \, ,
\end{equation}
where $\mathbb{I}$ is the $N\times N$ identity matrix (see Ref. \cite{ReviewJCN_2021} for a detailed description). The sets of left and right eigenvectors of $\mathbf{W}$ are the same for the normalized Laplacian  $\hat{\mathcal{L}}$. The respective eigenvalues $\xi_l$ of  $\hat{\mathcal{L}}$ are given by $\xi_l=1-\lambda_l$. 
\\[2mm]
In a similar way, we can deduce the eigenvalues $\xi_l^\prime$ and eigenvectors $\left|\phi^\prime_i\right\rangle$,  $\left\langle\bar{\phi}^\prime_l\right|$ of $\hat{\mathcal{L}}^\prime$, satisfying the conditions in Eqs. (\ref{cond1}) and (\ref{cond2}). Therefore, in terms of the sets of eigenvalues and eigenvectors of the matrices $\hat{\mathcal{L}}$, $\hat{\mathcal{L}}^\prime$, we have
\begin{eqnarray}
\exp[-\hat{\mathcal{L}}\,t]&=\sum_{l=1}^N\exp[-\xi_l\,t]\left|\phi_l\right\rangle\left\langle\bar{\phi}_l\right|,\\
\exp[-\hat{\mathcal{L}}^\prime\,t]&=\sum_{l=1}^N\exp[-\xi^\prime_l\,t]\left|\phi^\prime_l\right\rangle\left\langle\bar{\phi}^\prime_l\right|.
\end{eqnarray}
Then, by using the definitions in Eqs. (\ref{stateLleft})-(\ref{statesright}), we obtain
\begin{equation}\label{prod1}
    \langle \bar{\Psi}_i(t)| \Psi_i^\prime(t) \rangle=\sum_{l,m=1}^N e^{-(\xi_l+\xi^\prime_m)\,t} \left\langle i|\phi_l\right\rangle\left\langle\bar{\phi}_l|\phi^\prime_m\right\rangle\left\langle\bar{\phi}^\prime_m|i\right\rangle.
\end{equation}
In this relation, the values $\left\langle\bar{\phi}_l|\phi^\prime_m\right\rangle$ quantify the differences between the two bases associated to $\hat{\mathcal{L}}$, $\hat{\mathcal{L}}^\prime$. In a similar manner, we have
\begin{equation}\label{prod2}
    \langle \bar{\Psi}_i(t)| \Psi_i(t) \rangle=\sum_{l=1}^Ne^{-2\xi_l\,t}\left\langle i|\phi_l\right\rangle\left\langle\bar{\phi}_l|i\right\rangle,
\end{equation}
\begin{equation}\label{prod3}
    \langle \bar{\Psi}^\prime_i(t)| \Psi_i^\prime(t) \rangle=
    \sum_{l=1}^N e^{-2\xi^\prime_l\,t}\left\langle i|\phi^\prime_l\right\rangle\left\langle\bar{\phi}^\prime_l|i\right\rangle.
\end{equation}
Then, the introduction of Eqs. (\ref{prod1})-(\ref{prod3})  into Eq. (\ref{DissimilarM_nodei}) allows the calculation of  the  dissimilarity $\mathcal{D}_i(\hat{\mathcal{L}},\hat{\mathcal{L}}^\prime;t)$ considering the initial node $i$. In addition, it is convenient to calculate the global value
\begin{equation}\label{average_DissimilarM_nodes}
\bar{\mathcal{D}}(t)\equiv\frac{1}{N}\sum_{i=1}^N |\mathcal{D}_i(\hat{\mathcal{L}},\hat{\mathcal{L}}^\prime;t)|
\end{equation}
and the maximum global dissimilarity given by
\begin{equation}\label{MaxD_general}
|\bar{\mathcal{D}}|_{\mathrm{max}}\equiv\max\{\bar{\mathcal{D}}(t):t\geq 0\}.
\end{equation}
Finally, it is worth mentioning that in the temporal evolution given by Eq. (\ref{master}) for random walks defined in such a way that the walker can reach any node from any initial condition, the eigenvalue $\xi_1=0$ is unique. The eigenvector  associated to $\xi_1=0$ defines the stationary distribution $p_j^\infty(\hat{\mathcal{L}})$ that describes the probability for $t\to\infty$ and takes the form 
\begin{equation}
p_j^\infty(\hat{\mathcal{L}})=\left\langle i|\phi_1\right\rangle\left\langle\bar{\phi}_1|j\right\rangle.
\end{equation}
However, $\left\langle i|\phi_1\right\rangle$ is constant and, as a consequence, $p_j^\infty(\hat{\mathcal{L}})$ is independent of the initial condition. A similar result is valid for the dynamics with $\hat{\mathcal{L}}^\prime$
\begin{equation}
p_j^\infty(\hat{\mathcal{L}}^\prime)=\left\langle i|\phi^\prime_1\right\rangle\left\langle\bar{\phi}^\prime_1|j\right\rangle.
\end{equation}
Therefore, considering the limit $t\to\infty$ for $\mathcal{D}_i(\hat{\mathcal{L}},\hat{\mathcal{L}}^\prime;t)$ in Eq. (\ref{DissimilarM_nodei}), we have

\begin{eqnarray}
&\lim_{t\to \infty} \langle \bar{\Psi}_i(t)| \Psi_i(t) \rangle=\left\langle i|\phi_1\right\rangle\left\langle\bar{\phi}_1|i\right\rangle=p_i^\infty(\hat{\mathcal{L}}),\\
&\lim_{t\to \infty} \langle \bar{\Psi}^\prime_i(t)| \Psi_i^\prime(t) \rangle=\left\langle i|\phi^\prime_1\right\rangle\left\langle\bar{\phi}^\prime_1|i\right\rangle=p_i^\infty(\hat{\mathcal{L}}^\prime)
\end{eqnarray}
 and 
\begin{eqnarray}\nonumber
&\lim_{t\to \infty} \langle \bar{\Psi}_i(t)| \Psi_i^\prime(t) \rangle=\left\langle i|\phi_1\right\rangle\left\langle\bar{\phi}_1|\phi^\prime_1\right\rangle\left\langle\bar{\phi}^\prime_1|i\right\rangle\\\nonumber
&=\sum_{l=1}^N \left\langle i|\phi_1\right\rangle\left\langle\bar{\phi}_1|l\right\rangle \left\langle l|\phi_1^\prime\right\rangle\left\langle\bar{\phi}_1^\prime|i\right\rangle=\sum_{l=1}^N p_l^\infty(\hat{\mathcal{L}}) p_i^\infty(\hat{\mathcal{L}}^\prime),\\
&=p_i^\infty(\hat{\mathcal{L}}^\prime) 
\end{eqnarray}
to obtain
\begin{equation}\label{D_t_infty}
\mathcal{D}_i(\hat{\mathcal{L}},\hat{\mathcal{L}}^\prime;t\to \infty)=1-\frac{p_i^\infty(\hat{\mathcal{L}}^\prime)}{\sqrt{p_i^\infty(\hat{\mathcal{L}})\,p_i^\infty(\hat{\mathcal{L}}^\prime)}}.
\end{equation}
We see that $\mathcal{D}_i(\hat{\mathcal{L}},\hat{\mathcal{L}}^\prime;t\to \infty)$ is a comparison of the stationary probability distributions associated to the processes defined by $\hat{\mathcal{L}}$ and $\hat{\mathcal{L}}^\prime$. In particular, if the stationary distributions of these processes coincide, we have $\mathcal{D}_i(\hat{\mathcal{L}},\hat{\mathcal{L}}^\prime;t\to \infty)=0$.
\section{Dynamics on circulant networks}
\label{Section_Circulant}
In this section, we apply the general theory in Sec. \ref{Sec_General_Theory} to the study of processes defined in terms of circulant matrices. In this case, the matrices  $\hat{\mathcal{L}}$ and $\hat{\mathcal{L}}^\prime$ commute and have the same set of eigenvectors, allowing us to simplify $\mathcal{D}_i(\hat{\mathcal{L}},\hat{\mathcal{L}}^\prime;t)$ in  Eq. (\ref{DissimilarM_nodei}).  The results are illustrated with the analysis of the diffusion on interacting cycles and biased random walks on rings.
\subsection{Circulant matrices}
Let us now apply the general definition in Eq. (\ref{DissimilarM_nodei}) to the particular case where  $\hat{\mathcal{L}}$ and $\hat{\mathcal{L}}^\prime$ are defined by circulant matrices. A circulant matrix $\mathbf{C}$ is a  $N \times N$ matrix defined by \cite{VanMieghem2011}
\begin{equation} \label{matC}
\mathbf{C}=
\left(
\begin{array}{ccccc}
c_0 & c_{N-1} & c_{N-2} & \ldots & c_1\\
c_1 & c_{0} & c_{N-1} & \ldots & c_2\\
c_2 & c_{1} & c_{0} & \ldots & c_3\\
\vdots & \vdots & \vdots & \ddots & \vdots \\
c_{N-1} & c_{N-2} & c_{N-3} &\ldots & c_0\\
\end{array}\right) \, ,
\end{equation}
with entries denoted by $C_{ij}$. Then, each column has real elements $c_0, c_1,\ldots,c_{N-1}$ ordered in such a way that $ c_0 $ describes the diagonal elements and $C_{ij}=c_{(i-j)\textrm{mod}\, N}$. In cases where the network is a regular structure formed by cycles \cite{VanMieghem2011}, different types of random walks can be defined with a circulant matrix $\hat{\mathcal{L}}$; in particular, random walks with long-range displacements \cite{FractionalBook2019,RiascosMateosFD2015}, biased transport on directed rings  \cite{Biased_Pizarro2020}, among many others \cite{FractionalBook2019,Reset_2022}.
\\[2mm]
The elementary circulating matrix  $\mathbf{E}$ is defined with all its null elements except $c_1=1$. From $\mathbf{E}$, the integer powers $\mathbf{E}^l$ for $l=0,1,2,\ldots,N-1$ are also circulant matrices with null elements except $c_l=1$. Therefore, Eq. (\ref{matC}) can be expressed as \cite{VanMieghem2011}
\begin{equation}
\mathbf{C}=\sum_{m=0}^{N-1} c_m \mathbf{E}^{m}, \label{matCsum}
\end{equation}
where $\mathbb{I}=\mathbf{E}^{0}$. Furthermore, the relation $\mathbf{E}^N=\mathbb{I}$ requires that the eigenvalues  $\nu$ of $\mathbf{E}$  satisfy $\nu^N=1$; therefore, those eigenvalues are given by \cite{VanMieghem2011}
\begin{equation}
\nu_l=e^{\textrm{i}\frac{2\pi(l-1)}{N}} \qquad \textrm{for}\qquad l=1,\ldots, N,
\end{equation}
with $\textrm{i}\equiv\sqrt{-1}$. The respective eigenvectors $\{|\Psi_m\rangle\}_{m=1}^{N}$ have the components $\langle l |\Psi_m\rangle=\frac{1}{\sqrt{N}}e^{-\textrm{i}\frac{2\pi}{N}(l-1)(m-1)}$ and $\langle\Psi_l|m\rangle=\frac{1}{\sqrt{N}}e^{\textrm{i}\frac{2\pi}{N}(l-1)(m-1)}$ \cite{VanMieghem2011}. Now, considering Eq. (\ref{matCsum}), the eigenvectors $|\Psi_l\rangle$ satisfy $\mathbf{C}|\Psi_l\rangle=\eta_l |\Psi_l\rangle$, where the eigenvalues $\eta_l$ are given by (see Ref. \cite{VanMieghem2011} for details)
\begin{equation} \label{SpectCeta}
\eta_l=\sum_{m=0}^{N-1} c_m e^{\textrm{i}\frac{2\pi}{N}(l-1)\, m}
\end{equation}
for $l=1,2,\ldots, N$. All these analytical results show that the eigenvectors of circulant matrices are the same, whereas the respective eigenvalues include the values of the coefficients $c_0, c_1,\ldots,c_{N-1}$.
\\[2mm]
Then, if both matrices $\hat{\mathcal{L}}$ and $\hat{\mathcal{L}}^\prime$ are circulant  in the expressions in Eqs. (\ref{prod1})-(\ref{prod3}), we have $\left\langle\bar{\phi}_l|\phi^\prime_m\right\rangle=\delta_{lm}$, $\left\langle i|\phi_l\right\rangle\left\langle\bar{\phi}_l|i\right\rangle=\left\langle i|\phi^\prime_l\right\rangle\left\langle\bar{\phi}^\prime_l|i\right\rangle=1/N$. 
In addition
\begin{eqnarray} \nonumber
\hat{\mathcal{L}}\hat{\mathcal{L}}^\prime&=\sum_{l=1}^N\sum_{m=1}^N \xi_l\xi^\prime_m\left|\phi_l\right\rangle\left\langle\bar{\phi}_l\big|
\phi^\prime_m\right\rangle\left\langle\bar{\phi}^\prime_m\right|
\\\nonumber
&=\sum_{l=1}^N\xi_l\xi^\prime_l\left|\phi_l\right\rangle\left\langle\bar{\phi}_l\right|=\sum_{l=1}^N\xi_l\xi^\prime_l\left|\phi^\prime_l\right\rangle\left\langle\bar{\phi}^\prime_l\right|\\ \nonumber
&=\sum_{l=1}^N\sum_{m=1}^N\xi_l\xi^\prime_l\left|\phi^\prime_l\right\rangle\left\langle\bar{\phi}^\prime_l\big|
\phi_m\right\rangle\left\langle\bar{\phi}_m\right|\\ \nonumber
&=\sum_{l=1}^N\sum_{m=1}^N\xi_m\xi^\prime_l\left|\phi^\prime_l\right\rangle\left\langle\bar{\phi}^\prime_l\big|
\phi_m\right\rangle\left\langle\bar{\phi}_m\right|\\ \nonumber
&=\sum_{l=1}^N\xi^\prime_l\left|\phi^\prime_l\right\rangle\left\langle\bar{\phi}^\prime_l\right|\sum_{m=1}^N\xi_m\left|
\phi_m\right\rangle\left\langle\bar{\phi}_m\right|\\ 
&=\hat{\mathcal{L}}^\prime\hat{\mathcal{L}},
\end{eqnarray}
showing that in this case $\hat{\mathcal{L}}$ and $\hat{\mathcal{L}}^\prime$  commute. On the other hand
\begin{equation}\label{prod1_ciculant}
    \langle \bar{\Psi}_i(t)| \Psi_i^\prime(t) \rangle=\frac{1}{N}\sum_{l=1}^N e^{-(\xi_l+\xi^\prime_l)\,t},
\end{equation}
\begin{equation}\label{prod2_ciculant}
    \langle \bar{\Psi}_i(t)| \Psi_i(t) \rangle=\frac{1}{N}\sum_{l=1}^Ne^{-2\xi_l\,t},
\end{equation}
\begin{equation}\label{prod3_circulant}
    \langle \bar{\Psi}^\prime_i(t)| \Psi_i^\prime(t) \rangle=\frac{1}{N}
    \sum_{l=1}^N e^{-2\xi^\prime_l\,t}.
\end{equation}
As a consequence, Eq. (\ref{DissimilarM_nodei}) takes the form
\begin{equation}\label{DissimilarM_circulant}
\mathcal{D}(t)=1-\frac{\sum_{l=1}^{N}e^{-(\xi_{l}+\xi_{l}^\prime)t}}{\sqrt{\left|\sum_{l=1}^{N}e^{-2\xi_{l}t}\right|\left|\sum_{m=1}^{N}e^{-2\xi_{m}^\prime t}\right|}}
\end{equation}
where the eigenvalues $\xi_{l}$, $\xi_{l}^\prime$ are given by Eq. (\ref{SpectCeta}) considering the respective entries defining $\hat{\mathcal{L}}$ and $\hat{\mathcal{L}}^\prime$. In contrast with the general definition in Eq. (\ref{DissimilarM_nodei}), in Eq. (\ref{DissimilarM_circulant}) we see that $\mathcal{D}(t)$ is independent of the initial node $i$. This is because, when $\hat{\mathcal{L}}$ and $\hat{\mathcal{L}}^\prime$ are circulant matrices, the symmetry of the dynamics allows each node to be seen as equivalent. This is the case, for example, when we define a standard random walker in a regular network. 
\\[2mm]
As particular limits of $\mathcal{D}(t)$, we have $\mathcal{D}(0)=0$ and $\lim_{t\to\infty}\mathcal{D}(t)=0$. In addition, Eq. (\ref{average_DissimilarM_nodes}) takes the form $\bar{\mathcal{D}}(t)=|\mathcal{D}(t)|$; therefore, we have for  the maximum global dissimilarity in Eq. (\ref{MaxD_general})
\begin{equation}\label{MaxD_circulant}
|\mathcal{D}|_{\mathrm{max}}=\max\{|\mathcal{D}(t)|: t\geq 0\}.
\end{equation}
In the following, we evaluate two particular cases in which the values of $\mathcal{D}(t)$ and $|\mathcal{D}|_{\mathrm{max}}$ are studied.
\subsection{Diffusion on interacting cycles}
\label{Sec_Intec_Cycles}
\begin{figure*}[!t]
	\begin{center}
		\includegraphics*[width=1.0\textwidth]{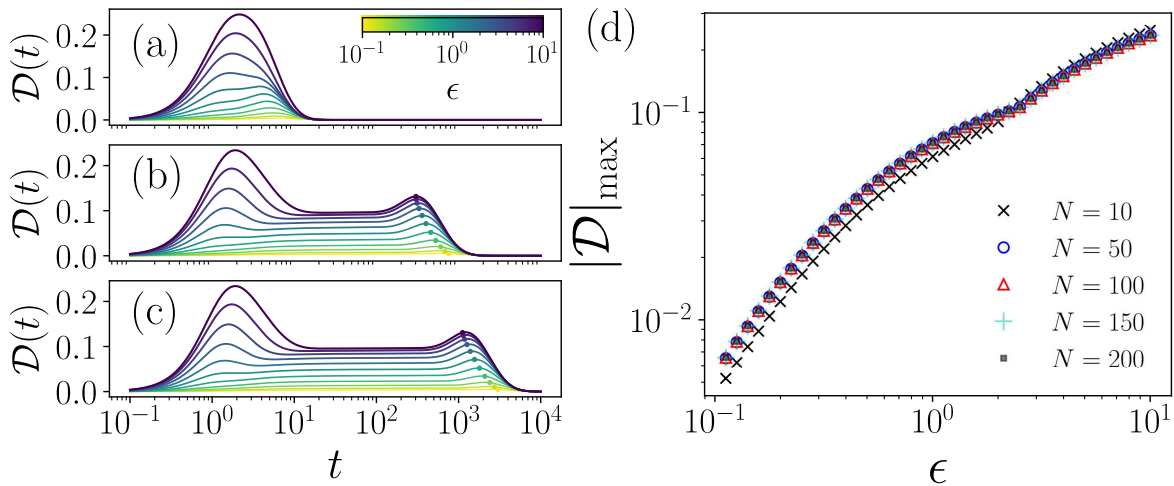}
	\end{center}
	\vspace{-5mm}
	\caption{\label{Fig_1} Dissimilarity between the standard diffusion on a ring and on weighted interacting cycles. $\mathcal{D}(t)$ as a function of $t$ for networks with sizes: (a) $N=10$, (b) $N=100$, (c) $N=200$, the weights $\epsilon$ are codified in the colorbar. The results are obtained with the numerical evaluation of Eq. (\ref{DissimilarM_circulant}) with the eigenvalues (\ref{eigen_ring}) and (\ref{eigen_interacting_cycle}). In panels (b)-(c) we represent with dots the value $\mathcal{D}(t^\star)$ for $t^\star=\mathcal{K}/1.75$. (d) $|\mathcal{D}|_{\mathrm{max}}$ in Eq. (\ref{MaxD_circulant}) as a function of $\epsilon$ for networks with different sizes $N$.}
\end{figure*}

In this section, we explore the dynamics of a standard random walker with transition probabilities defined by $w_{i\to j}=A_{ij}/k_i$ \cite{ReviewJCN_2021,NohRieger2004}. The random walker hops between nodes with equal probability from a node to one of its neighbors. For the dynamics on a finite ring with $N$ nodes, the modified Laplacian $\hat{\mathcal{L}}$ is a circulant matrix defined by $c_0=1$, $c_1=-1/2$, $c_{N-1}=-1/2$, therefore, through the application of Eq. (\ref{SpectCeta}) we have the eigenvalues of $\hat{\mathcal{L}}$
\begin{equation}\label{eigen_ring}
\xi_{l}=1-\cos\left[\frac{2\pi}{N}(l-1)\right].
\end{equation}
For the dynamics with $\hat{\mathcal{L}}^\prime$, we consider a modification of the initial ring with $N$ nodes to add a set of edges with a weight $\epsilon>0$ linking each node with nodes at distance $2$ in the original ring. The new edges are an extension of the local neighborhood with interactions weighted with the parameter $\epsilon$. In this structure, a random walker moving through the links and considering the weights is defined by a circulant matrix $\hat{\mathcal{L}}^\prime$ with non-null elements $c_0=1$, $c_1=c_{N-1}=-1/(2+2\epsilon)$, $c_2=c_{N-2}=-\epsilon/(2+2\epsilon)$. Therefore, the application of Eq. (\ref{SpectCeta}) allows to deduce the eigenvalues for $\hat{\mathcal{L}}^\prime$ given by
\begin{equation}\label{eigen_interacting_cycle}
\xi^\prime_{l}=1-\frac{1}{1+\epsilon}\cos\left[\frac{2\pi}{N}(l-1)\right]
-\frac{\epsilon}{1+\epsilon}\cos\left[\frac{4\pi}{N}(l-1)\right]
\end{equation}
with $\epsilon\geq 0$. In particular, the limit $\epsilon\to 0$ recovers the result in Eq. (\ref{eigen_ring}) for a ring. Then, from the analytical results for the eigenvalues in Eqs. (\ref{eigen_ring}) and (\ref{eigen_interacting_cycle}), we can evaluate the dissimilarity $\mathcal{D}(t)$ in Eq. (\ref{DissimilarM_circulant}) between the dynamics on the ring and the case with weighted links. The results are reported in Fig. \ref{Fig_1} for different values of $\epsilon$ and sizes $N$. 
\\[2mm]
In panels \ref{Fig_1}(a)-(c), we present $\mathcal{D}(t)$ as function of $t$ using $0.1\leq \epsilon\leq 10$, the numerical results are shown with curves with $\epsilon$ codified in the colorbar for networks with $N=10,\, 100, \, 200$. If $\epsilon\ll 1$, the results show that $\mathcal{D}(t)$ is relatively small for all $t$. On the other hand, for $\epsilon>0$, $\mathcal{D}(t)$ presents two relative maximums, this is clear for the cases with $\epsilon\gg 1$ and $N$ large [Figs. \ref{Fig_1}(b)-(c)]; however, for $N=10$ [Fig. \ref{Fig_1}(a)] the two peaks overlap. In the results, a first maximum is found at small times $t\approx 2$ because the two random walk dynamics initially differ due to the effect of the weighted links with $\epsilon$. However, this modification is local, the global effect at large times produces a second maximum, this is a consequence of modifications at large scale in the connectivity for $N$ large. In panels Fig. \ref{Fig_1}(b)-(c) we show that the second maximum lies around the value $t^\star=\mathcal{K}/1.75$ where $\mathcal{K}$ is the Kemeny's constant \cite{ReviewJCN_2021,Kemeny}
\begin{equation*}
\mathcal{K}=\sum_{l=2}^N \frac{1}{\xi^\prime_l}
\end{equation*}
that in regular networks quantifies the average number of steps to reach any node considering all the initial conditions (see Ref. \cite{ReviewJCN_2021} for a detailed discussion).    
\\[2mm]
For $t\to \infty$, $\mathcal{D}(t)\to 0$ since the two processes have the same stationary probability $p_j^\infty(\hat{\mathcal{L}})=p_j^\infty(\hat{\mathcal{L}}^\prime)=1/N$.
\\[2mm]
Once analyzed  $\mathcal{D}(t)$, in Fig.  \ref{Fig_1}(d) we depict  $|\mathcal{D}|_{\mathrm{max}}$ that gives the maximum $|\mathcal{D}(t)|$ for $t>0$. We represent  $|\mathcal{D}|_{\mathrm{max}}$  as a function of $\epsilon$ for $N=10,50,100,150,200$. Here, it is worth to noticing that the values of $|\mathcal{D}|_{\mathrm{max}}$ are similar for networks with $N$ large ($N\geq 50$) and some small deviations appear for the case with $N=10$. In addition, the results show that $|\mathcal{D}|_{\mathrm{max}}$ increases monotonically with $\epsilon$ evidencing the gradual differences generated with the introduction of the new set of weighted edges.
\subsection{Biased transport on rings}
\begin{figure*}[!t]
	\begin{center}
		\includegraphics*[width=1.0\textwidth]{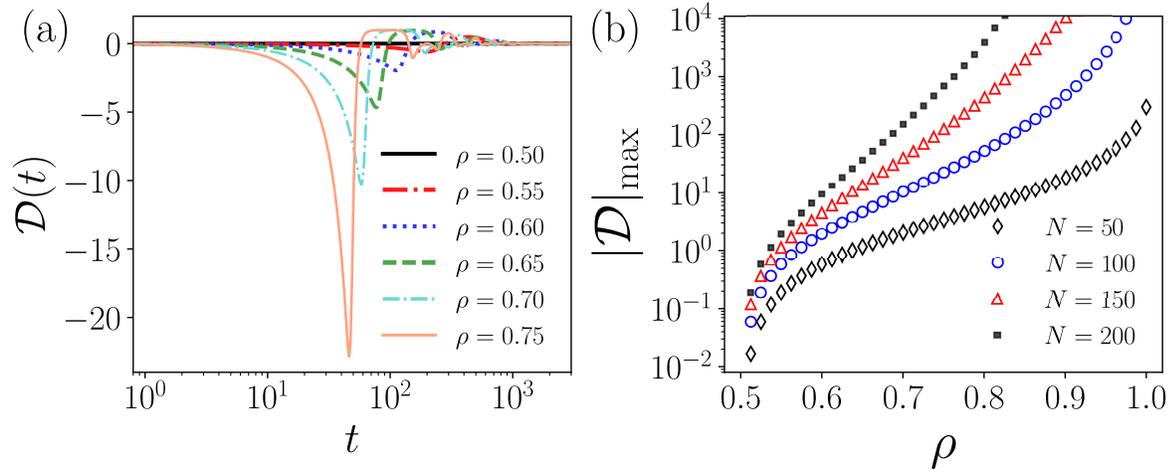}
	\end{center}
	\vspace{-5mm}
	\caption{\label{Fig_2} Dissimilarity between the standard diffusion and the biased transport on rings. (a) $\mathcal{D}(t)$ as a function of $t$ for different bias $\rho$.  The results are obtained with the numerical evaluation of Eq. (\ref{DissimilarM_circulant}) with the eigenvalues in Eqs. (\ref{eigen_ring}) and (\ref{eigen_ring_biased2}) for networks with $N=100$.  (b) $|\mathcal{D}|_{\mathrm{max}}$ in Eq. (\ref{MaxD_circulant}) as a function of $\rho$ for rings with sizes $N=50,100,150,200$. }
\end{figure*}
A second example for the dissimilarity between dynamical processes defined with circulant matrices is obtained when we compare the standard random walk on a ring with a matrix $\hat{\mathcal{L}}$
with non-null elements $c_0=1$, $c_{1}=c_{N-1}=-1/2$ and eigenvalues given in Eq. (\ref{eigen_ring}) with the biased transport on a ring defined by a modified Laplacian  $\hat{\mathcal{L}}^\prime$ with non-null elements $c_0=1$, $c_1=-(1-\rho)$, $c_{N-1}=-\rho$ with $0\leq \rho\leq 1$. In this case, $\rho$ and $1-\rho$ define the probabilities to pass from one node to one of the two neighbors on a ring (on a circular layout, at each node the walker chooses to move clockwise with probability $\rho$ and counterclockwise with probability $1-\rho$).  Therefore, the application of Eq. (\ref{SpectCeta}) leads to the eigenvalues of $\hat{\mathcal{L}}^\prime$
\begin{equation}\label{eigen_ring_biased}
\xi^\prime_{l}=1-(1-\rho)\exp\left[\textrm{i}\frac{2\pi}{N}(l-1)\right]
-\rho\exp\left[-\textrm{i}\frac{2\pi}{N}(l-1)\right].
\end{equation}
The unbiased dynamics defined by $\hat{\mathcal{L}}$ is recovered when $\rho=1/2$. In particular, using $\varphi_l\equiv\frac{2\pi}{N}(l-1)$, we can rewrite Eq. (\ref{eigen_ring_biased}) to have
\begin{equation}\label{eigen_ring_biased2}
\xi^\prime_{l}=\xi_l+\textrm{i}(2\rho-1)\sin\varphi_l.
\end{equation}
The value $\Delta_l=(2\rho-1)\sin\varphi_l$ quantifies the modifications in the eigenvalues $\xi_l$ of $\hat{\mathcal{L}}$ in Eq. (\ref{eigen_ring}). In particular, the asymmetry of the transport produces complex eigenvalues generating an oscillatory behavior in the values of $\mathcal{D}(t)$. In Eq. (\ref{DissimilarM_circulant}) the two sums including information of $\hat{\mathcal{L}}^\prime$ take the form
\begin{eqnarray}
\sum_{l=1}^{N}e^{-(\xi_{l}+\xi_{l}^\prime)t}&=\sum_{l=1}^{N}e^{-2\xi_{l}t}\cos[(2\rho-1)t\sin\varphi_l],\\
\sum_{l=1}^{N}e^{-2\xi_{l}^\prime t}&=\sum_{l=1}^{N}e^{-2\xi_{l}t}\cos[2(2\rho-1)t\sin\varphi_l].
\end{eqnarray}
In Fig. \ref{Fig_2} we explore the numerical values of $\mathcal{D}(t)$ and $|\mathcal{D}|_{\mathrm{max}}$ for  $0.5\leq \rho<1$. In Fig. \ref{Fig_2}(a), we show $\mathcal{D}(t)$ as a function of $t$ for $N=100$ and $\rho=0.5,\, 0.55,\ldots,\,0.75$. The results illustrate the oscillatory behavior of  $\mathcal{D}(t)$ for different values of $\rho$. For $\rho=1/2$, $\mathcal{D}(t)=0$.
In Fig. \ref{Fig_2}(b), we depict the maximum value of $|\mathcal{D}(t)|$ as a function of $\rho$ for rings with sizes $N=50,100,150, 200$. The results show how $|\mathcal{D}|_{\mathrm{max}}$ is affected by the size of the network $N$ and the bias $\rho$, with a completely different behavior to the observed with the addition of weighted links in Fig. \ref{Fig_1}(d). We see that in Fig. \ref{Fig_2}(b), a comparison of the unbiased dynamics with the biased dynamics with $\rho>0.6$ produces an important modification of the transport evidenced in large values of  $|\mathcal{D}|_{\mathrm{max}}$.  The results for $|\mathcal{D}|_{\mathrm{max}}$ increase monotonically in the interval $0.5< \rho<1$ and for large rings $|\mathcal{D}|_{\mathrm{max}}\gg 1$ for $\rho>0.9$ showing that in these cases the dynamics differ significantly when compared with the unbiased transport. In addition, the bias alters globally the exploration of the rings reducing the average time necessary to reach any node of the network \cite{Biased_Pizarro2020}. In particular, $\rho=1$ produces a deterministic case where the walker moves only clockwise (or only counterclockwise) visiting one of the nearest neighbors at each step requiring $N$ steps to visit all the nodes, this deterministic limit is completely different from the random dynamics generated with $\rho=1/2$.
\section{Characterizing the  effect of rewiring}
\label{Sec_rewiring}
In the cases aforementioned for circulant matrices in Sec. \ref{Section_Circulant},  all the nodes suffered the same modifications in the second process defined by $\hat{\mathcal{L}}^\prime$. In this manner, $\mathcal{D}(t)$ in Eq. (\ref{DissimilarM_circulant}) captures the differences between the two dynamical processes analyzed. However, other modifications of the Laplacian $\hat{\mathcal{L}}^\prime$ may be due to heterogeneous alterations in the nodes. In this section, we explore two cases where we compare the effect of introducing a new edge in a ring and the differences generated by random rewiring in a regular network.
\subsection{Ring with an additional link}
\begin{figure*}[!t]
	\begin{center}
		\includegraphics*[width=1.0\textwidth]{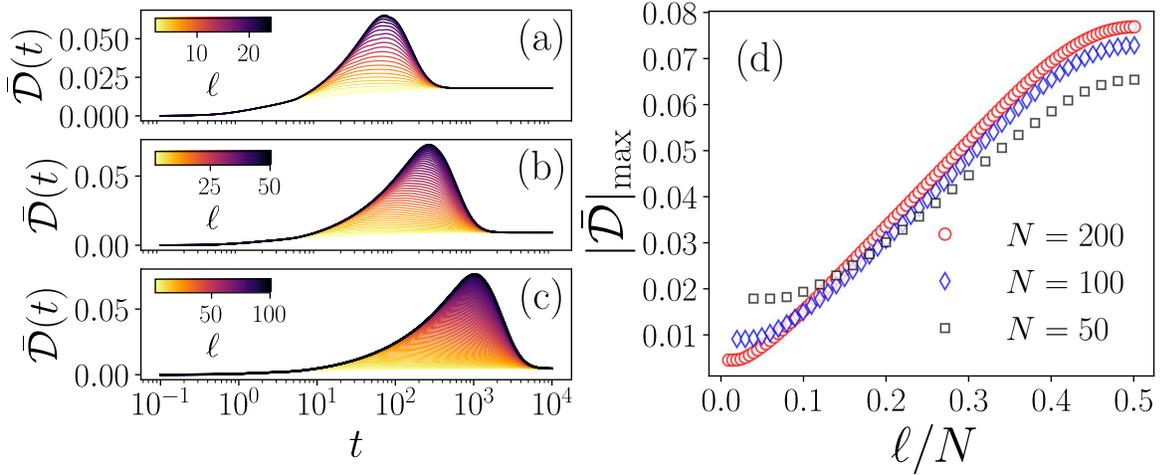}
	\end{center}
	\vspace{-5mm}
	\caption{\label{Fig_3} Dissimilarity between the standard diffusion on a ring and on a ring with an additional edge. $\bar{\mathcal{D}}(t)$ as a function of $t$ for different values of $\ell=2,3,\ldots,\lfloor N/2\rfloor$, codified in the colorbar. The results are obtained with the numerical evaluation of Eqs. (\ref{DissimilarM_nodei}) and (\ref{average_DissimilarM_nodes}) for networks with: (a) $N=50$, (b) $N=100$, (c) $N=200$. (d)  $|\bar{\mathcal{D}}|_{\mathrm{max}}$ in Eq. (\ref{MaxD_general}) as a function of $\ell/N$ for networks with $N=50,100,200$. }
\end{figure*}

In this case, we compare the dynamics generated by $\hat{\mathcal{L}}$ for a standard random walk on a ring with $N$ nodes as defined in Sec. \ref{Sec_Intec_Cycles}. The second process generated by $\hat{\mathcal{L}}^\prime$ describes the same dynamics but on a different network defined by a ring with an additional edge connecting two nodes at a distance $\ell=2,3,\ldots,\lfloor N/2\rfloor$ in the original ring. In the graph theory literature this type of edge is called a chord \cite{west2001graph_theory}.
\\[2mm]
In the following, we explore the effect of $\ell$ when we compare the diffusive dynamics in the ring and the ring with a chord. Here, it is worth mentioning that in all the cases the matrices $\hat{\mathcal{L}}$ and $\hat{\mathcal{L}}^\prime$ differ in particular entries. However, for any $\ell$, we have $\sum_{i,j}|\mathcal{L}_{ij}-\mathcal{L}^\prime_{ij}|=4/3$, showing that the direct comparison of the matrical elements does not capture the differences between the dynamics generated by $\hat{\mathcal{L}}$ and  $\hat{\mathcal{L}}^\prime$.
\\[2mm]
Therefore, for a comparison of $\hat{\mathcal{L}}$ and  $\hat{\mathcal{L}}^\prime$ in the context of diffusion, we use the average value $\bar{\mathcal{D}}(t)$ in Eq. (\ref{average_DissimilarM_nodes}) and its maximum value $|\bar{\mathcal{D}}|_{\mathrm{max}}$ in Eq. (\ref{MaxD_general}). The numerical results for networks  with $N=50,\,100,\,200$ are shown in Fig. \ref{Fig_3}.
In panels \ref{Fig_3}(a)-(c) we depict  $\bar{\mathcal{D}}(t)$ as a function of $t$, the results are presented as different curves generated for $\ell=2,3,\ldots,\lfloor N/2\rfloor$. The numerical values of $\bar{\mathcal{D}}(t)$ show that $\bar{\mathcal{D}}(t)\approx 0 $ for $t$ small, then $\bar{\mathcal{D}}(t)$ gradually increases; in particular, for $\ell=2$, the increments are monotonic until reaching a plateau. For  $\ell\gg 2$ we observe a peak that rises with $\ell$. The results are in good agreement with the fact that introducing a chord with small $\ell$, the averages of $|\mathcal{D}_i(\hat{\mathcal{L}},\hat{\mathcal{L}}^\prime;t)|$ over all the nodes $i$ are small since the chord only produces little variations affecting the global dynamics, these differences reduce when we increase the size of the network $N$. Furthermore, $\ell$ large creates greater connectivity that substantially changes the dynamics with respect to the original ring. The results show that for $\ell$ large, the time $t$ where is produced the maximum of $\bar{\mathcal{D}}(t)$  increases with the size of the network.
\\[2mm]
On the other hand, in the limit $t\to \infty$, $\bar{\mathcal{D}}(t)$ shows a comparison between the two stationary distributions of the random walk dynamics generated with $\hat{\mathcal{L}}$,  $\hat{\mathcal{L}}^\prime$. Using Eq. (\ref{D_t_infty})
\begin{eqnarray}\nonumber
&\bar{\mathcal{D}}(\infty)\equiv\lim_{t\to\infty}\bar{\mathcal{D}}(t)\\\nonumber
&= \frac{(N-2)}{N}\left|1-\sqrt{\frac{N}{N+1}}\right|+\frac{2}{N}\left|1-\sqrt{\frac{3N}{2(N+1)}}\right|\\
&=1-\frac{4}{N}-
\sqrt{\frac{N}{N+1}}+\frac{\sqrt{6}+2}{\sqrt{N(N+1)}}  ,
\end{eqnarray}
with the asymptotic expansion for  $N$ large 
\begin{equation}
\bar{\mathcal{D}}(\infty)=\frac{\sqrt{6}-3/2}{N}+O(N^{-2}).
\end{equation}
Therefore, $\bar{\mathcal{D}}(\infty)$ decreases with $N$ for $N\geq 3$. In particular, for $N=50,100, 200$, $\bar{\mathcal{D}}(\infty)$ takes the values $0.01797$, $0.009237$, $0.004683$, respectively. These results agree with the numerical values obtained for $t$ large in Figs. \ref{Fig_3}(a)-(c).
\\[2mm]
In Fig. \ref{Fig_3}(d) we present the values of the maximum dissimilarity $|\bar{\mathcal{D}}|_{\mathrm{max}}$ in Eq. (\ref{MaxD_general}) in terms of the value $\ell/N$. The results allow us to compare the dynamics on the ring and the effect of the chord. For the different sizes of the networks, we see $|\bar{\mathcal{D}}|_{\mathrm{max}}$ increases monotonically with $\ell$.
\subsection{Rewiring  using the Watts-Strogatz model }
Let us now consider the effect of random rewiring on networks. To this end, we define $\hat{\mathcal{L}}$ describing the standard random walk on a regular network and $\hat{\mathcal{L}}^\prime$ for the same dynamics on a new structure generated with the stochastic rewiring of multiple links. The networks are obtained with the Watts-Strogatz model \cite{WattsStrogatz1998}. In this case, a random network is generated as follows:  a ring is connected to the same number $J$ of nearest neighbors on each side, $2J$ is the degree of each node. This network resembles a one-dimensional lattice with periodic boundary conditions. Then, 
a Watts–Strogatz network is created by removing each edge with uniform, independent probability $p$ and rewiring it to connect a pair of nodes that are chosen uniformly at random \cite{WattsStrogatz1998}. In the Watts–Strogatz model, the rewiring procedure modifies the global connectivity of the network inducing the small-world property. This topological feature is characterized by the mean shortest path length \cite{Barrat2000}
\begin{figure*}[!t]
	\begin{center}
		\includegraphics*[width=1.0\textwidth]{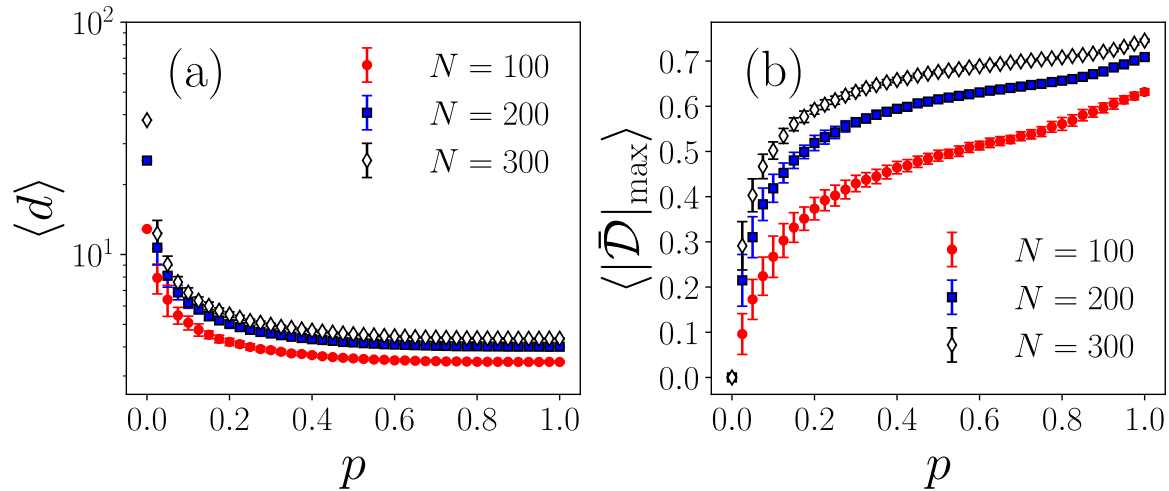}
	\end{center}
	\vspace{-5mm}
	\caption{\label{Fig_4} Comparison of the diffusion on a regular network and Watts-Strogatz random networks with rewiring probability $p$. (a)  Ensemble average $\langle d \rangle$ of the mean path length in Eq. (\ref{meanspl}). (b) Ensemble average $\langle|\bar{\mathcal{D}}|_{\mathrm{max}}\rangle$ of the maximum dissimilarity in Eq. (\ref{MaxD_general}). We consider 100 realizations of the random networks generated for each value of $p$, the error bars represent the standard deviation of the results, the sizes of the network are $N=100,\, 200,\, 300$.}
\end{figure*}
\begin{equation}\label{meanspl}
d\equiv \frac{1}{N(N-1)}\sum_{i\neq j} d_{ij},
\end{equation}
where $d_{ij}$ is the length of the shortest path connecting the nodes $i$ and $j$. 
\\[2mm]
We analyze the effect of rewiring for networks described by $J=2$ and the probability $p$ comparing a random walker in the structure obtained for $p=0$ and the dynamics when $0\leq p\leq 1$. For $p=0$, the network is regular and $\hat{\mathcal{L}}$ is a circulant matrix defined by non-null elements $c_0=1$, $c_{1}=c_{2}=c_{N-1}=c_{N-2}=-1/4$. The second process defined by $\hat{\mathcal{L}}^\prime$ takes place on the modified network generated with rewiring $p$ and transition probabilities  $w_{i\to j}=A_{ij}/k_i$. 
\\[2mm]
In Fig. \ref{Fig_4}, we explore the mean path length $d$ in (\ref{meanspl}) and the maximum dissimilarity $|\bar{\mathcal{D}}|_{\mathrm{max}}$ in Eq. (\ref{MaxD_general}). The results are obtained numerically for networks with $N=100,\, 200,\, 300$ and different values of $p$. Since, for each $p$ the rewiring produces a different network, we evaluate the results for $100$ realizations, we present the ensemble average $\langle d \rangle$ and  $\langle|\bar{\mathcal{D}}|_{\mathrm{max}}\rangle$ over realizations, the error bars show the respective standard deviation.
\\[2mm]
The values of  $\langle d \rangle$ help us to understand the effect of the rewiring in the network topology.  In Fig. \ref{Fig_4}(a) we depict $\langle d \rangle$ as a function of $p$. The results show that for $p=0$ average distances between nodes are maximum; however, the rewiring generates a long-range connectivity in the structure reducing $\langle d \rangle$ with $p$. We see that, for $p>0.6$ the variations of $\langle d \rangle$ are relatively small in comparison with the changes observed for $0<p\leq 0.2$.
\\[2mm]
In Fig. \ref{Fig_4}(b) we explore $\langle|\bar{\mathcal{D}}|_{\mathrm{max}}\rangle$ as a function of $p$ to compare the random walker in a modified network with the dynamics on the original structure (with $p=0$). The results show that for a fixed $p$, 
$\langle|\bar{\mathcal{D}}|_{\mathrm{max}}\rangle$ increases with the size $N$, this relation is reasonable since although $p$ is the same, the number of rewired links increase with $N$ producing larger differences with the transport on the regular network for the dynamics with $\hat{\mathcal{L}}$. On the other hand, for networks with the same size, the values of  $\langle|\bar{\mathcal{D}}|_{\mathrm{max}}\rangle$ increase monotonically with $p$ showing that the dynamics in the modified network differs gradually with the rewiring generated by $p$. However, the rate of increase is greater in the interval $0<p<0.2$ showing that a little fraction of rewiring produces significant changes in the diffusion. 
\section{Comparing different random walk strategies}
\label{Sec_DifferentRWs}
In the applications described in Secs. \ref{Section_Circulant} and \ref{Sec_rewiring}, we compared the dynamics of random walkers to see the effect of modifications in the network through the introduction of new edges, changes in weights, and rewiring. In the following part, we describe two cases where the network is the same but the random walker is modified with a particular parameter.
\subsection{Degree-biased random walks}
In this section, we discuss degree-biased random walks. For this case, the random walker hops with transition probabilities $w_{i \to j}$ depending on the degrees of the neighbors of the node $i$. Degree-biased random walks are defined by a transition matrix $\mathbf{W}(\beta)$ with elements \cite{FronczakPRE2009} 
\begin{equation}\label{wijBRWglobal}
	w_{i\rightarrow j}(\beta)=\frac{A_{ij} k_j^{\beta}}{\sum_{l=1}^N A_{il} k_l^{\beta}},
\end{equation}
where $\beta$ is a real parameter. In Eq.~(\ref{wijBRWglobal}), $\beta>0$ describes the bias to hop to neighbor nodes with a higher degree, whereas for $\beta<0$ this behavior is inverted and, the walker tends to hop to nodes less connected. When $\beta=0$, the normal random walk with $w_{i\to j}=A_{ij}/k_i$ is recovered.
\\[2mm]
The random walk in Eq. (\ref{wijBRWglobal}) is also defined in terms of a symmetric matrix of weights $\mathbf{\Omega}$ with elements $\Omega_{ij}=A_{ij}(k_i k_j)^\beta$ with transition probabilities \cite{ReviewJCN_2021}
\begin{equation*}
    w_{i\rightarrow j}=\frac{\Omega_{ij}}{\sum_{l=1}^N \Omega_{il}}=\frac{\Omega_{ij}}{S_i}.
\end{equation*}
Here $S_i=\sum_{l=1}^N \Omega_{il}$ and represents the total weight of the node $i$ (see Ref. \cite{ReviewJCN_2021} for a review of different random walks defined using a symmetric matrix of weights). In terms of this formalism, in connected undirected networks, degree-biased random walks are ergodic for $\beta$ finite and the stationary distribution is given by \cite{ReviewJCN_2021}
\begin{equation}\label{StatPBRW}
	p_i^{\infty}=\frac{S_i}{\sum_{l=1}^N S_l}=\frac{\sum_{l=1}^N (k_i k_l)^\beta A_{il}}{\sum_{l,m=1}^N (k_l k_m)^\beta A_{lm}} \, .
\end{equation}
Degree  biased random walks have been studied extensively in the literature in different contexts as varied as routing processes \cite{WangPRE2006}, chemical reactions \cite{KwonPRE2010}, extreme events \cite{KishorePRE2012,LingEPJB2013}, multiple random walks on networks \cite{WengPRE2017,RiascosSandersPRE2021}, among others \cite{FronczakPRE2009,LambiottePRE2011,Battiston2016}. Recently, degree-biased random walks have been generalized to include multiple biases \cite{Wang_Chaos2021}, in potential-driven random walks  \cite{BenigniPRE2021}, to examine the influence of damage and aging in complex systems  \cite{Aging_PhysRevE2019,Eraso_Hernandez_2021} and, to incorporate a particular bias in each node \cite{Riascos_LocalBiasPRE2022}.
\\[2mm]
Once defined the transition matrix for degree-biased random walks, we can use this information in the definition of a normalized Laplacian matrix associated to the continuous-time dynamics 
\begin{equation}\label{normLbiased_ref}
\hat{\mathcal{L}}(\beta)=\mathbb{I}-\mathbf{W}(\beta).
\end{equation}
Here $\beta$ is a given value of the bias parameter that serves as a reference to compare it with a second process
\begin{equation}\label{normLbiased_prime}
\hat{\mathcal{L}}^\prime(\beta^\prime)=\mathbb{I}-\mathbf{W}(\beta^\prime).
\end{equation}
In this manner, we can compare the dynamics with bias parameter $\beta^\prime$  with a particular reference given by $\beta$.
\\[2mm]
\begin{figure*}[!t]
	\begin{center}
		\includegraphics*[width=1.0\textwidth]{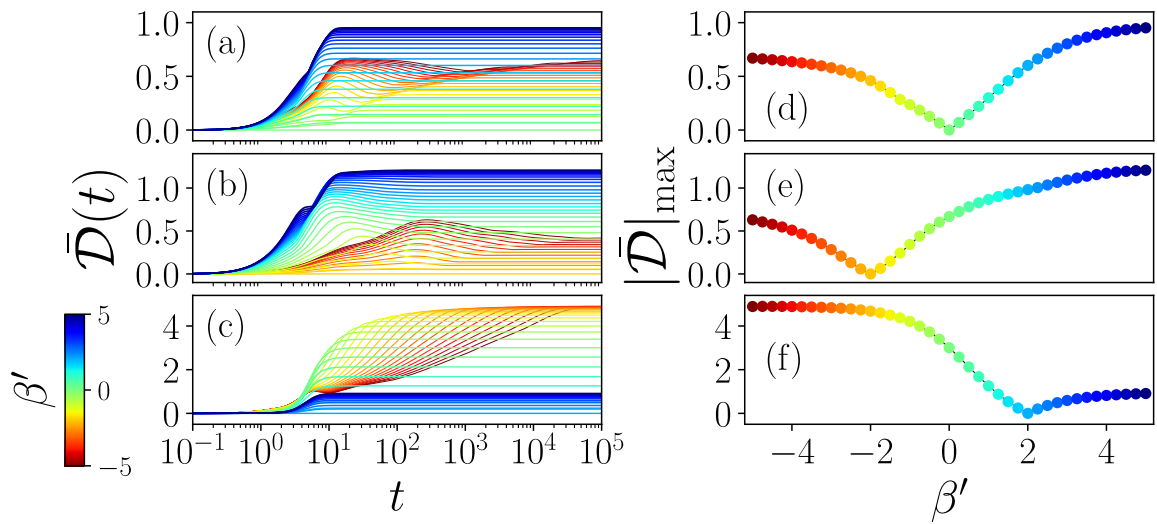}
	\end{center}
	\vspace{-7mm}
	\caption{\label{Fig_5} Comparison between degree-biased random walks on a network of the  Barab\'asi-Albert  type with $N=100$ nodes and $m=2$. $\bar{\mathcal{D}}(t)$ as a function of $t$ for modified processes with $-5\leq\beta^\prime\leq 5$ (codified in the colorbar) using as reference a dynamics defined by: (a) $\beta=0$, (b) $\beta=-2$, (c) $\beta=2$. Panels (d)-(f) depict the maximum dissimilarity  $|\bar{\mathcal{D}}|_{\mathrm{max}}$ as a function of $\beta^\prime$ for references (d) $\beta=0$, (e) $\beta=-2$, (f) $\beta=2$ (markers are colored using $\beta^\prime$ as in the left panels). }
\end{figure*}
In Fig. \ref{Fig_5}, we calculate numerically $\bar{\mathcal{D}}(t)$ in Eq. (\ref{average_DissimilarM_nodes}) and   $|\bar{\mathcal{D}}|_{\mathrm{max}}$ in Eq. (\ref{MaxD_general}) as a function of $\beta^\prime$ considering three particular values of the reference value $\beta$ for degree-biased random walks on a  Barab\'asi-Albert network. This structure is generated randomly by adding new nodes each with $m$ edges that are preferentially attached to existing nodes with higher degree \cite{BarabasiAlbert1999}. The degrees present different values with the existence of some hubs and a high fraction of nodes with few neighbors. Due to this heterogeneity, the degree-biased random walk strategy differs from the unbiased case. In the analysis presented in Fig. \ref{Fig_5}, the size of the network is $N=100$ and was generated using $m=2$. In Figs.  \ref{Fig_5}(a)-(c) we compare the degree-biased random walk using $\beta=0,-2,2$ with a modified dynamics defined by $-5\leq \beta^\prime\leq 5$. We plot $\bar{\mathcal{D}}(t)$ as a function of $t$, the values of $\beta^\prime$ are codified in the colorbar. The maximum dissimilarities $|\bar{\mathcal{D}}|_{\mathrm{max}}$ as a function of  $\beta^\prime$ are presented in Figs.  \ref{Fig_5}(d)-(f) for  $\beta=0$ in panel (d), $\beta=-2$ in (e) and $\beta=2$ in  (f). As a guide to the eye, we maintained the same color code for  $\beta^\prime$ as in panels \ref{Fig_5}(a)-(c).
\\[2mm]
Our analysis in Fig. \ref{Fig_5} for $\bar{\mathcal{D}}(t)$ reveal how the average differences between two processes evolve; it also allows the identification of the times in which its maximum values occur and shows us how $\bar{\mathcal{D}}(t)$  reaches a stationary value for large times. In particular,  the limit $\bar{\mathcal{D}}(t\to\infty)$ can be obtained analytically introducing into Eq. (\ref{D_t_infty})  the stationary distributions given by Eq. (\ref{StatPBRW}). On the other hand, $|\bar{\mathcal{D}}|_{\mathrm{max}}$ gives the maximum average dissimilarity. This value characterizes the differences between the two dynamics defined by $\beta$ and $\beta^\prime$ without the details of the time $t$ where this maximum occurs. In particular, we see $|\bar{\mathcal{D}}|_{\mathrm{max}}=0$ when the process used as reference and the modified dynamics coincide, i.e. when $\beta=\beta^\prime$. For $\beta\neq\beta^\prime$, the results show that $|\bar{\mathcal{D}}|_{\mathrm{max}}$ increases as $|\beta-\beta^\prime|$ increases, this is valid in both intervals $\beta>\beta^\prime$ and $\beta<\beta^\prime$. It is worth to observe the asymmetry of $|\bar{\mathcal{D}}|_{\mathrm{max}}$ around the reference value $\beta$. For example, in  Fig. \ref{Fig_5}(f) where $\beta=2$, the maximum dissimilarity $|\bar{\mathcal{D}}|_{\mathrm{max}}$ remains relatively small for  $\beta^\prime>2$ in comparison with the values for $\beta^\prime<0$ evidencing marked differences between the respective processes.
\\[2mm]
The numerical values reported in Fig. \ref{Fig_5} also show that the dissimilarity measure implemented in Eq. (\ref{DissimilarM_nodei}) depends on the order of the reference $\hat{\mathcal{L}}$ and the second process $\hat{\mathcal{L}}^\prime$ as we discussed in Sec. \ref{Subsec_Ds}. For example, in Fig. \ref{Fig_5}(d) for $\beta=0$, $\beta^\prime=2$ we have  $|\bar{\mathcal{D}}|_{\mathrm{max}}=0.602$ whereas in Fig. \ref{Fig_5}(f) for $\beta=2$, $\beta^\prime=0$,  $|\bar{\mathcal{D}}|_{\mathrm{max}}=3.006$.

\subsection{Random walks with reset}
When a stochastic process is interrupted and restarted from the initial state, its occupation probability in the configuration space is strongly altered \cite{Evans_2020}. Interestingly, the mean time needed to reach a given target state for the first time can often be minimized with respect to the resetting rate \cite{Bonomo2021}. In this section, we consider a random walk with stochastic reset to a particular node \cite{ResetNetworks_PRE2020}. The walker performs two types of steps: with probability $1-\gamma$, a random jump from the node currently occupied to a different node of the network, or, with probability $\gamma$, a resetting to a fixed node $r$. Without resetting ($\gamma=0$), the probability to hop to $m$ from $l$ is $w_{l\to m}$, the random walk is ergodic and described by the transition matrix $\mathbf{W}$.
\\[2mm]
The dynamics with stochastic reset is defined by the transition probability matrix $\mathbf{\Pi}(r;\gamma)$ with elements 
\begin{equation}\label{Transition_Reset}
\pi_{l \to m}(r;\gamma)\equiv (1-\gamma) w_{l\to m}+\gamma\,\delta_{rm}, 
\end{equation}
where $\sum_{m=1}^N \pi_{l \to m}(r;\gamma)=1$. The matrix
$\mathbf{\Pi}(r;\gamma)$ completely entails the process with resetting, which is able to reach all the nodes of the network if the resetting probability $\gamma$ is $<1$. The matrices $\mathbf{W}$ and $\mathbf{\Pi}(r;\gamma)$ are stochastic matrices: knowing their eigenvalues and eigenvectors allows the calculation of the occupation probability at any time, including the stationary distribution at $t\to\infty$, as well as the mean first passage time to any node. The eigenvalues and eigenvectors of $\mathbf{\Pi}(r;\gamma)$  are related to those of $\mathbf{W}$, which is recovered in the limit $\gamma=0$ (see a detailed discussion in Refs. \cite{ResetNetworks_PRE2020,Reset_2022,Touchette_PRE2018,MultipleResetPRE_2021}).
\\[2mm]
The connection between the eigenvalues $\lambda_l$ of $\mathbf{W}$ and $\zeta_l(r;\gamma)$ for $\mathbf{\Pi}(r;\gamma)$ is obtained from the relation \cite{ResetNetworks_PRE2020}
\begin{equation}\label{Def_MatPi_R1}
\mathbf{\Pi}(r;\gamma)=(1-\gamma)\mathbf{W}+\gamma \mathbf{\Theta}(r),
\end{equation}
where the elements of the matrix $\mathbf{\Theta}(r)$ are $\Theta_{lm}(r)=\delta_{mr}$. Namely, $\mathbf{\Theta}(r)$ has entries $1$ in the $r^{th}$-column and null entries everywhere else, therefore (see Ref. \cite{ResetNetworks_PRE2020} for details)

\begin{equation}\label{eigvals_zeta}
\zeta_l(r;\gamma)=
\left\{\begin{array}{c@{\quad}l} 
1  &\mathrm{for}\qquad l=1,\\
(1-\gamma)\lambda_l  &\mathrm{for}\qquad l=2,3,\ldots, N.
\end{array}\right.
\end{equation}
This result reveals that the eigenvalues are independent of the choice of the resetting node $r$. On the other hand, the left eigenvectors of $\mathbf{\Pi}(r;\gamma)$ are given by  \cite{ResetNetworks_PRE2020}
\begin{equation}\label{psil1}
\left\langle\bar{\psi}_1(r;\gamma)\right|=\left\langle\bar{\phi}_1\right|
+\sum_{m=2}^N\frac{\gamma}{1-(1-\gamma)\lambda_m}\frac{\left\langle r|\phi_m\right\rangle}{\left\langle r|\phi_1\right\rangle}\left\langle\bar{\phi}_m\right|
\end{equation}
whereas $\left\langle\bar{\psi}_l(r;\gamma)\right|=\left\langle\bar{\phi}_l\right|$ for $l=2,\ldots,N$. Similarly, the right eigenvectors are given by:
$\left|\psi_1(r;\gamma)\right\rangle=\left|\phi_1\right\rangle$ and \cite{ResetNetworks_PRE2020}
\begin{equation}\label{psirl_reset}
\left|\psi_l(r;\gamma)\right\rangle=
\left|\phi_l\right\rangle-\frac{\gamma}{1-(1-\gamma)\lambda_l}\frac{\left\langle r|\phi_l\right\rangle }{\left\langle r|\phi_1\right\rangle}\left|\phi_1\right\rangle,
\end{equation}
for $l=2,\ldots,N$.
\\[2mm]
Once described the dynamics with resetting and modifications that the stochastic restart introduces in the eigenvalues and eigenvectors of $\mathbf{\Pi}(r;\gamma)$, let us now to compare the process defined in continuous time by the matrix 
\begin{equation}\label{Laplacian_noreset}
\hat{\mathcal{L}}=\mathbb{I}-\mathbf{W}
\end{equation}
with eigenvalues $\xi_l=1-\lambda_l$ and right and left eigenvectors $\left|\phi_l\right\rangle$, $\left\langle \bar{\phi}_l\right|$ and a second process
 with resetting given by
\begin{equation}\label{Laplacian_reset}
\hat{\mathcal{L}}^\prime=\mathbb{I}-\mathbf{\Pi}(r;\gamma)
\end{equation}
with eigenvalues $\xi^\prime_l=1-\zeta_l(r;\gamma)$ where $\zeta_l(r;\gamma)$ is given by Eq. (\ref{eigvals_zeta}) and eigenvectors $\left|\psi_l(r;\gamma)\right\rangle$, $\left\langle \bar{\psi}_l(r;\gamma)\right|$ in Eqs. (\ref{psil1})-(\ref{psirl_reset}).
\\[2mm]
Here, it is important to notice that using the definitions in Eqs. (\ref{Laplacian_noreset}) and (\ref{Laplacian_reset}): $\hat{\mathcal{L}}-\hat{\mathcal{L}}^\prime=\gamma\left[\mathbf{\Theta}(r)-\mathbf{W}\right]$. Therefore the direct comparison of the elements of $\hat{\mathcal{L}}$ and $\hat{\mathcal{L}}^\prime$ gives
\begin{equation}\label{Diff_Lap_Reset}
\sum_{i,j=1}^N|\mathcal{L}_{ij}-\mathcal{L}_{ij}^\prime|=\gamma\sum_{i,j=1}^N|\delta_{rj}-w_{i\to j}|.
\end{equation}
Showing that this particular comparison of elements is proportional to $\gamma$ and does not give more information of the differences between the two dynamics.
\\[2mm]
For the processes defined by $\hat{\mathcal{L}}$ and $\hat{\mathcal{L}}^\prime$, we evaluate the relations in Eqs. (\ref{prod1})-(\ref{prod3}) to express the results in terms of $\left|\phi_l\right\rangle$, $\left\langle \bar{\phi}_l\right|$ for the dynamics without resetting. We obtain
\begin{eqnarray}\nonumber
\langle \bar{\Psi}_i(t)&|\Psi_i^\prime(t)\rangle= 
\sum_{l=1}^N\langle i|\phi_l\rangle\langle \bar{\phi}_l|i\rangle e^{-(\xi_l+\xi^\prime_l)t}\\
&+\gamma\sum_{m=2}^N\frac{1}{\xi^\prime_m }\langle r|\phi_m\rangle\langle \bar{\phi}_m|i\rangle\left(1-e^{-\xi^\prime_m t}\right),
\end{eqnarray}
\begin{equation}
\langle \bar{\Psi}_i(t)|\Psi_i(t)\rangle= 
\sum_{l=1}^N\langle i|\phi_l\rangle\langle \bar{\phi}_l|i\rangle e^{-2\xi_l t},
\end{equation}
\begin{eqnarray}\nonumber
\langle \bar{\Psi}^\prime_i(t)&|\Psi^\prime_i(t)\rangle= 
\sum_{l=1}^N\langle i|\phi_l\rangle\langle \bar{\phi}_l|i\rangle e^{-2\xi^\prime_l t}\\
&+\gamma\sum_{m=2}^N\frac{1}{\xi^\prime_m }\langle r|\phi_m\rangle\langle \bar{\phi}_m|i\rangle\left(1-e^{-2\xi^\prime_m t}\right).
\end{eqnarray}
Using this information, we can evaluate $\mathcal{D}_i(\hat{\mathcal{L}},\hat{\mathcal{L}}^\prime;t)$ in Eq. (\ref{DissimilarM_nodei}) to obtain the maximum dissimilarity $|\bar{\mathcal{D}}|_{\mathrm{max}}$ in Eq. (\ref{MaxD_general}). 
\\[2mm]
\begin{figure*}[!t]
	\begin{center}
		\includegraphics*[width=1.0\textwidth]{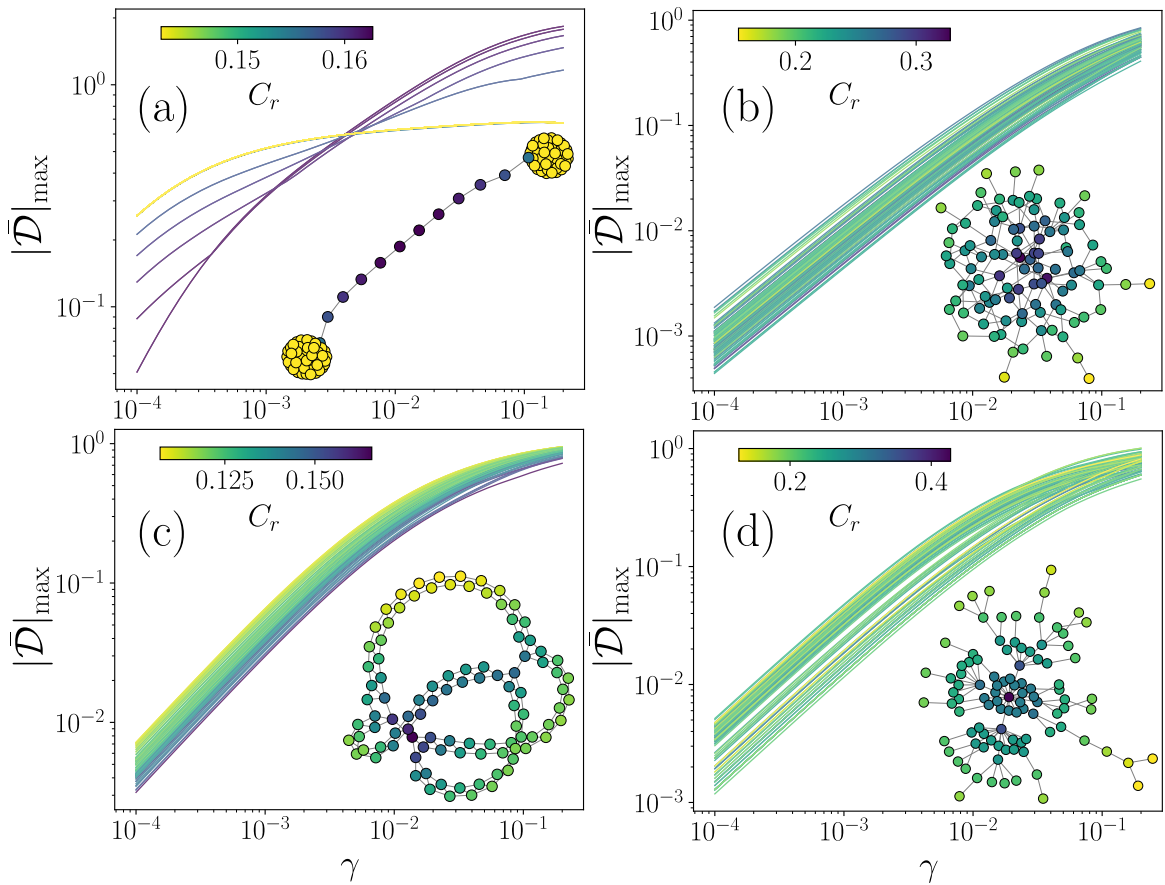}
	\end{center}
	\vspace{-5mm}
	\caption{\label{Fig_6}  Maximum dissimilarity  $|\bar{\mathcal{D}}|_{\mathrm{max}}$ between standard random walks and random dynamics with reset to the node $r$ with probability $\gamma$ for networks with $N=100$ nodes:  (a) Barbell, (b) Watts-Strogatz, (c) Erd\"{o}s-R\'enyi and, (d) Barab\'asi-Albert, where each newly introduced node connects to $m$ previous nodes (we use $m=1$ in the generation of the network). We depict the values of  $|\bar{\mathcal{D}}|_{\mathrm{max}}$ as a function of $\gamma$ for all the nodes $r=1,\ldots,N$. To identify the effects of resetting, we colored each node $r$ and the curves with results using respective closeness centrality $C_r \equiv \frac{N}{\sum_{j=1}^N d_{rj}}$ of node $r=1,\ldots,N$.}
\end{figure*}
In Fig. \ref{Fig_6}, we present the numerical values of $|\bar{\mathcal{D}}|_{\mathrm{max}}$ as a function of $\gamma$ for networks with $N=100$ considering all the values of $r$ for the resetting node, each network is presented as an inset. The node $r$ and the curves with results are colored the using respective closeness centrality $C_r \equiv \frac{N}{\sum_{j=1}^N d_{rj}}$ of node $r=1,\ldots,N$ ($d_{rj}$ is the length of the shortest path connecting the nodes $r$ and $j$).  In Fig. \ref{Fig_6}(a) we analyze a Barbell graph (constructed by connecting two fully connected networks with $45$ nodes with a line of $10$ nodes) \cite{Saberi_SIAM_2008}. In Fig. \ref{Fig_6}(b) a Watts-Strogatz network \cite{WattsStrogatz1998} with rewiring probability $p=0.01 $. In panel \ref{Fig_6}(c) an Erd\"{o}s-R\'enyi network  \cite{ErdosRenyi1959} with average degree $\langle k \rangle= 2.72$, and in \ref{Fig_6}(d)  a Barab\'asi-Albert network generated with the preferential attachment rule with $m=1$ \cite{BarabasiAlbert1999}.
\\[2mm]
In general, our findings in Fig. \ref{Fig_6} show the global difference between the standard diffusion and the diffusion with stochastic reset on networks. The characterization given by  $|\bar{\mathcal{D}}|_{\mathrm{max}}$ goes beyond what can be obtained by a direct comparison of the matrical elements in Eq. (\ref{Diff_Lap_Reset}) revealing the dependence with the node $r$ where the reset is produced. For the Barbell graph in panel (a), $|\bar{\mathcal{D}}|_{\mathrm{max}}$ shows major differences in the dynamics generated by the nodes $r$ where the random walk is reset. In particular, all the nodes in the fully connected subgraphs (with the lowest $C_r$) produce the same curve for $|\bar{\mathcal{D}}|_{\mathrm{max}}$ as a function of $\gamma$ with $\gamma$ producing slow variations of $|\bar{\mathcal{D}}|_{\mathrm{max}}$ for  $\gamma\geq 0.004$. In contrast, for the same interval, the variations of $|\bar{\mathcal{D}}|_{\mathrm{max}}$ increase with $C_r$ when the reset is produced to the nodes in the linear subgraph. In the random networks generated with the Erd\"{o}s-R\'enyi and Barab\'asi-Albert algorithms (panels (b) and (d)), the values $|\bar{\mathcal{D}}|_{\mathrm{max}}$ have similar behavior for all the nodes, and in both cases the effect of $C_r$ is not clear. On the contrary, for the  Watts-Strogatz network (panel (c)), the curves describing $|\bar{\mathcal{D}}|_{\mathrm{max}}$ move upward when resetting is produced to nodes with lower $C_r$. 

\section{Conclusions}
We explore the comparison between diffusive processes on networks. The implemented measure quantifies the dissimilarity of states that evolve with the information consigned in normalized Laplacians $\hat{\mathcal{L}}$ and $\hat{\mathcal{L}}^\prime$ for two dynamical processes. We use the dissimilarity $\mathcal{D}_i(\hat{\mathcal{L}},\hat{\mathcal{L}}^\prime;t)$ defined in terms of eigenvalues and eigenvectors of $\hat{\mathcal{L}}$ and $\hat{\mathcal{L}}^\prime$ to compare the states at time $t$ of the systems considering the initial node $i$. In particular,  $\mathcal{D}_i(\hat{\mathcal{L}},\hat{\mathcal{L}}^\prime;t=0)=0$ and  $\mathcal{D}_i(\hat{\mathcal{L}},\hat{\mathcal{L}}^\prime;t\to\infty)$ establishes a comparison of the respective stationary distributions. A global quantity $\bar{\mathcal{D}}(t)$ and  its maximum $|\bar{\mathcal{D}}|_{\mathrm{max}}$ are also introduced.
\\[2mm]
We illustrated all the mathematical framework implemented with the exploration of several cases as follows: First, we analyze dynamics where $\hat{\mathcal{L}}$ and $\hat{\mathcal{L}}^\prime$ are circulant matrices. In this case, the comparison between the two processes depends exclusively on the eigenvalues of each matrix. Secondly, we explore the effect of the addition of a new edge in a ring and the effect of stochastic rewiring using the Watts-Strogatz model. In a third case, we compare the transport when the network is the same but the way a random walker hops between nodes changes. We explored degree-biased random walks and local random walks with stochastic reset to the initial node.
\\[2mm]
For all the dynamics explored, we observed that the quantities analyzed provide a method to compare matrices defining diffusive dynamics on networks considering the evolution of states and capturing the complexity of the diffusion on these structures. The methods do not rely on particularities of the operators associated with the diffusion and a similar approach using other dynamical processes can be useful to evaluate the effect of modifications in a complex system; for example, the introduction of a new route in a public transportation system, the reduction of the functionality in a system due to aging, the information spreading on temporal networks, among many others.

\section*{Acknowledgments}
APR and acknowledges support from Ciencia de Frontera 2019 (CONACYT), project ``Sistemas complejos estoc\'asticos: Agentes m\'oviles, difusi\'on de part\'iculas, y din\'amica de espines'' (Grant No. 10872). %
\section*{References}

\providecommand{\noopsort}[1]{}\providecommand{\singleletter}[1]{#1}%
\providecommand{\newblock}{}

\end{document}